\documentclass[11pt]{article}
\usepackage[a4paper]{geometry}

\usepackage{amsmath} %% Standard AMS stuff
\usepackage{amssymb} %% Standard AMS stuff
\usepackage{amsthm} %% Standard AMS stuff
\usepackage{bbm} %% Improved Blackboard Bold: \mathbbm
\usepackage{empheq}
\usepackage[loose]{subfigure} %% Subfigures: \subfigure inside figure environment
\usepackage{epsfig} %% Allow Encapsulated PostScript figures
\usepackage{graphicx}
\usepackage{psfrag}
\usepackage[usenames,dvipsnames]{pstricks}
\usepackage{pst-plot}
\usepackage[colorlinks]{hyperref} %% Hyperlinks
\usepackage{tabls} %% Better spacing around mathematics in tables
\usepackage{paralist}

\newcommand{\COMMENT}[1]{}
\newcommand{\E}{\mathbb{E}}

\renewcommand{\P}{\mathbb{P}}

\newcommand{\numeq}{\stackrel{\mathrm{num}}{=}}

\theoremstyle{definition}

\hypersetup{
    linkcolor=blue,          % color of internal links
}

\title{
\textbf{The optimal uncertainty algorithm in the \emph{mystic} framework} \\
}

\author{
M.\ McKerns\thanks{Materials Science, California Institute of Technology, Pasadena, CA 91125, USA.  Email:  \textsf{mmckerns@caltech.edu}}
, H.\ Owhadi\thanks{Applied \& Computational Mathematics, California Institute of Technology, Pasadena, CA 91125, USA.  Email:  \textsf{owhadi@caltech.edu}}
, C.\ Scovel\thanks{Modeling, Algorithms, \& Informatics, Los Alamos National Laboratory, Los Alamos, NM 87545, USA.  Email:  \textsf{jcs@lanl.gov}}
, T.\ J.\ Sullivan\thanks{Graduate Aerospace Laboratories, California Institute of Technology, Pasadena, CA 91125, USA.  Email:  \textsf{tjs@caltech.edu}}
,~\& M.\ Ortiz\thanks{Graduate Aerospace Laboratories, California Institute of Technology, Pasadena, CA 91125, USA.  Email:  \textsf{ortiz@aero.caltech.edu}}
}

\date{August 21, 2010}

\makeatletter
\@addtoreset{equation}{section}
\@addtoreset{figure}{section}
\@addtoreset{table}{section}
\makeatother

\begin{document}

\maketitle
%% REVISION HISTORY
%% 2010-08-16: branched from the OUQ paper body

\begin{abstract}
% We present an implementation of the optimal uncertainty quantification (OUQ) algorithm \cite{??} in the context of the \emph{mystic} optimization framework.
 We have recently proposed a rigorous framework for Uncertainty Quantification (UQ)
 in which UQ objectives and  assumption/information set are brought into the forefront, providing a framework for the communication and comparison of UQ results. In particular,  this framework   does not implicitly impose  
inappropriate assumptions nor does it repudiate  relevant information.

This framework, which we call \emph{Optimal Uncertainty Quantification} (OUQ), is based on the observation that given a set of assumptions and information, there exist bounds on uncertainties obtained as values of optimization problems and that these bounds are optimal.
 It provides a uniform environment for the optimal solution of the problems of
validation, certification, experimental design,      reduced order modeling, prediction, extrapolation, all  under aleatoric and epistemic uncertainties.

OUQ optimization problems are extremely large, and even though under general conditions they have finite-dimensional reductions, they must often be solved numerically.  This general algorithmic framework for OUQ has been implemented in the \emph{mystic} optimization framework. We describe this implementation, and demonstrate its use in the context of the Caltech surrogate model for hypervelocity impact.
\end{abstract}

\newpage %%!!!
%% REVISION HISTORY
%% 2010-08-16: branched from the OUQ paper body

\section{The OUQ algorithm in the \emph{mystic} framework}\label{sec:Mystic}

The \emph{mystic} optimization framework \cite{McKernsHungAivazis:2009} provides a collection of optimization algorithms and tools that lowers the barrier to solving complex optimization problems.  \emph{Mystic} provides a selection of optimizers, both global and local, including several gradient solvers.  The optimizers included in \emph{mystic} all use a common interface, so they can be easily interchanged without the user having to write new code.

Constraints in \emph{mystic} are classified as either ``bounds constraints'' (linear inequality constraints that involve precisely one input variable) or ``non-bounds constraints'' (constraints between two or more parameters).  Every \emph{mystic} optimizer provides the ability to apply bounds constraints generically and directly to the cost function, so that the difference in the speed of bounds-constrained optimization and unconstrained optimization is minimal.  \emph{Mystic} also enables the user to impose an arbitrary parameter constraint function on the input of the cost function --- allowing non-bounds constraints to be generically applied in any optimization problem.

In addition to standard penalty methods that couple the solution of the constraints with the solution of the optimization problem, \emph{mystic} provides methods for imposing constraints on a discrete set. Both the statistical methods and the constraints solvers in \emph{mystic} are thus built to decouple the constraints from the optimization problem. This implementation greatly reduces the complexity of the optimization problem, and the number of function evaluations does not blow up with the complexity of constraints.  Since evaluation of the model is commonly the most expensive part of the optimization, \emph{mystic} optimizers should also converge much more quickly than other algorithms that apply constraints by invalidating generated results (\emph{i.e.}\ filtering) at each iteration.  In this way, \emph{mystic} can efficiently solve for rare events because the set of input variables produced by the optimizer at each iteration will also be an admissible point in problem space --- this feature is critical in solving OUQ problems, as tens of thousands of function evaluations may be required to produce a solution.  
Our implementation of the OUQ algorithm in \emph{mystic} utilizes a nested optimization (\emph{i.e.}\ the OUQ inner loop) to solve an arbitrary set of parameter constraints at each evaluation of the cost function.  Hence, the OUQ algorithm depends on reliably obtaining accurate solutions to the constraints at each iteration.  Fortunately, \emph{mystic} provides solver-independent termination conditions --- a capability that greatly increases the flexibility for numerically solving problems with non-standard convergence profiles.

The implemention of OUQ within the \emph{mystic} framework is presented below.

\paragraph{OUQ as an optimization problem.} OUQ problems can be thought of optimization problems where the goal is to find the global maximum of a probability function $\mu[H = 0]$, where $H = 0$ is a failure criterion for the model response function $H$.  Additional conditions in an OUQ problem are provided as constraints on the information set. Typically, a condition such as a mean constraint on $H$, $m_{1} \leq \mathbb{E}_{\mu}[H] \leq m_{2}$, will be imposed on the maximization.
After casting the OUQ problem in terms of optimization and constraints, we can plug these terms into the infrastructure provided by \emph{mystic}.

In the most general sense, optimization is a feedback loop between a measure of ``goodness of fit'' (or \verb!cost!) for a chosen model and the optimization algorithm (see Figure \ref{fig:OptimizerLoop}).  The \verb!costFunction! is a measure of quality that evaluates the \verb!cost! for a set of model parameters \verb!params!. The natural metric for $\P[H = 0]$ is the optimization criteria, hence we select the evaluation of $\mu[H = 0]$ as the \verb!costFunction!.
The optimizer will generate trial parameters every iteration in the feedback loop until the termination conditions are met. Termination is typically decided by convergence criteria on the last n values of \verb!cost! and \verb!params!, however it can also be determined by conditions on the current values of \verb!cost! and \verb!params!, the number of \verb!costFunction! evaluations, the number of iterations of the feedback loop, or any combinations thereof.  Upon satisfying the termination conditions, the optimizer produces the calculated maximal $\mu[H = 0]$ as \verb!opt_cost!  and the calculated maximizers as \verb!opt_params!.

\COMMENT{
\begin{verbatim}
PICTURE OF THE FEEDBACK LOOP GOES HERE.

   ---->[costFunction]---->
  |                       |
params                  cost
  |                       |
  |---->-- params ---->   |
  |                   |   |
  <---[paramGen]--[terminate?]---> (opt_param, opt_cost)

\end{verbatim}
}
% Clint suggests the above figure be modified to include a box for "history".
% Essentially, params and cost feed into a logger, and the logger feeds the
% history into the termination checker.  I'm not sure if this is too conceptually
% difficult to add to the figure.  It's already in the text, and not adding the
% logger to the figure only implies that history is logged in the termination checker.
\begin{figure}[tp]
	\begin{center}
		%\includegraphics[width=0.45\textwidth]{./fig/optimizer_loop.eps}
		%% Mike's EPS figure replaced by PSTtricks code by TJS on 2010-07-26
		\begin{pspicture}(-1.0,-1.0)(9.0,5.0)
			\psframe[linewidth=0.04,framearc=0.2,dimen=outer](2.0,4.0)(6.0,5.0)
			\rput(4.0,4.5){\texttt{costFunction}}
			\psframe[linewidth=0.04,framearc=0.2,dimen=outer](1.0,1.0)(5.0,2.0)
			\rput(3.0,1.5){\texttt{paramGen}}
			\pscurve[linewidth=0.04,arrowsize=0.2 2.0,arrowlength=1.4,arrowinset=0.4]{->}(1.0,1.5)(0.5,2.0)(0.0,3.0)(1.5,4.25)(2.0,4.5)
			\rput(0.25,4.25){\texttt{params}}
			\psdiamond[linewidth=0.04,dimen=outer](6.5,1.5)(0.5,0.5)
			\rput[l](6.9,1.1){\text{Terminate?}}
			\pscurve[linewidth=0.04,arrowsize=0.2 2.0,arrowlength=1.4,arrowinset=0.4]{->}(6.0,4.5)(8.0,3.0)(7.0,1.5)
			\rput(7.5,4.25){\texttt{cost}}
			\pscurve[linewidth=0.04,arrowsize=0.2 2.0,arrowlength=1.4,arrowinset=0.4]{->}(1.0,1.5)(0.5,2.5)(1.5,3.0)(5.0,3.0)(6.5,2.0)
			\rput(4.0,3.5){\texttt{params}}
			\psline[linewidth=0.04,arrowsize=0.2 2.0,arrowlength=1.4,arrowinset=0.4]{->}(6.0,1.5)(5.0,1.5)
			\rput(5.5,1.8){\text{No}}
			\psline[linewidth=0.04,arrowsize=0.2 2.0,arrowlength=1.4,arrowinset=0.4]{->}(6.5,1.0)(6.5,0.0)
			\rput[r](6.25,0.5){\text{Yes}}
			\psframe[linewidth=0.04,framearc=0.2,dimen=outer](4.0,-1.0)(9.0,0.0)
			\rput(6.5,-0.5){\texttt{(opt\_cost, opt\_params)}}
		\end{pspicture}
		\caption{Conceptual diagram of an optimization loop. Parameters \texttt{params} are generated by the optimizer's parameter generator \texttt{paramGen} each iteration, and are piped into the cost function \texttt{costFunction}.  The cost function is evaluated at the model parameters and returns the \texttt{cost}.  The convergence criteria are checked after every iteration, and if the termination conditions are met, the optimizer stops. \texttt{opt\_cost} is the resulting calculated maximum, and \texttt{opt\_params} are the calculated maximizers.}
		\label{fig:OptimizerLoop}
	\end{center}
\end{figure}

\paragraph{Measures as data objects.} It is noted that a solution to an OUQ problem can be expressed in terms of product measures \cite{OUQ:2010}.  Thus, in a computational implementation of OUQ, the natural means for passing information between different elements of the algorithm code should also be based on product measures.
A hierarchy of parameterized measure data objects is the natural commodity for the evaluation of $\mu[H = 0]$. The optimizer's parameter generator \verb!paramGen! will produce new parameters each iteration, and hence produce new product measures to be evaluated within the \verb!costFuntion!.
For instance, a response function $H$ that requires input of dimension $n=3$, here defined by $H(x,y,z)$, will require a product measure of dimension $n=3$ for support. 
In problems with bounds constraints on the mean of the response function, such as $[m_{1},m_{2}] = [0,1]$, where the parameters $x,y,z$ may also be bounded by range, we then can use products of convex combinations of Dirac masses as the basis for support.
The corresponding OUQ code parameterizes the Dirac masses by their weights and positions. These weights and positions become the inputs for the optimization problem, and the solved weights and positions are thus the maximizers for the optimization problem.

The following general formulation is applicable to product measures of any dimension, and also for measures that utilize a basis other than Dirac masses.
However, for simplicity the following is described in the context of an optimization over Dirac masses of dimension $n=3$.
\vspace{\baselineskip}

\noindent We use the Dirac measure class, available from \emph{mystic}, to build a dimension $n=3$ product measure data object hierarchy that supports the transfer of information through the optimization loop (see Figure \ref{fig:OptimizerLoop}), where:
\begin{compactitem}
	\item a 3D product measure is composed of $3$ one-dimensional discrete measures --- this construction is the computational representation of the tensorization of measures to form a product measure;
	\item these one-dimensional discrete measures are data objects composed of $N$ ``support points'' --- these objects represent (not-necessarily-convex) linear combinations of weighted Dirac measures;
	\item support points are data objects that have mass and position --- these objects represent weighted Dirac measures on the one-dimensional input parameter spaces.
\end{compactitem}
\vspace{\baselineskip}

\noindent Discrete measures are provided with the methods:
\begin{compactitem}
	\item ``coords'': set or return the positions of each support point;
	\item ``weights'': set or return the mass of each support point;
	\item ``npts'': return the number of support points;
	\item ``mass'': calculate the sum of all weights;
	\item ``normalize'': set the sum of all weights to 1.0;
	\item ``range'': set or calculate ``$\max - \min$'' for the positions;
	\item ``mean'': set or calculate the centre of mass of the measure, \emph{i.e.}\ the Euclidean dot product of the ``coords'' array with the ``weights'' array.
\end{compactitem}
\vspace{\baselineskip}
As mentioned above, discrete measures are composed of finitely many support points, with each support point having weight and position. The first three of the above methods are required to describe the composition of the measure itself, while the last four methods provide the measure with additional mathematical properties.
It is worth noting that the implementation of ``normalize'', ``range'', and ``mean'' are such that setting one of these three properties does not alter the value of the other two.  For example, imposing a mean is done by translating the positions of all basis elements in a discrete measure equally, thereby preserving the measure's range and mass.  This implementation is an essential part of helping \emph{mystic} decouple constraints from the optimization problem.
Additionally, constraints on the range of the input parameters are applied directly to the cost function using the \verb!function_wrapper! method provided by \emph{mystic}. These parameter range constraints ensure that the weights are in the interval $[0, 1]$, while an additional constraints function enforces the normalization (total measure $= 1$) condition by calling ``normalize''.
\vspace{\baselineskip}

\noindent Product measures are provided with the methods:
\begin{compactitem}
	\item ``coords'': set or return the positions of each support point
	\item ``weights'': return the mass of each support point
	\item ``npts'': return the number of support points
	\item ``expect'': set or calculate $\E_{\mu}[H]$ for response function $H$
\end{compactitem}
\vspace{\baselineskip}
While product measures require the same three methods to describe the composition of the measure as discrete measures do, product measures are provided with one new property, the method ``expect'', which sets or calculates the expectation value.
A OUQ algorithm implementation requiring a mean constraint on $H$ thus defines a constraint function that ensures that $m_{1} \leq E_{\mu}[H] \leq m_{2}$, and additionally that the sum of the weights on each discrete measure is equal to 1.0.  Additional constraints on the information set are added to the optimization problem by adding conditions within the constraints function implemented using the above methods.
\vspace{\baselineskip}

\noindent Also, several helper utilities are available to aid in conversion between the different data object representations:
\begin{compactitem}
	\item ``unpack'': maps a $n$ dimensional product measure to its $n$ factors, which are 1D discrete measures;
	\item ``pack'': map $n$ 1D discrete measures to one dimension $n$ product measure, \emph{i.e.}\ forms their product measure;
	\item ``flatten'': map $n$ x 1D discrete measures to an input parameter list;
	\item ``unflatten'': map an input parameter list to $n$ x 1D discrete measures.
\end{compactitem}
\vspace{\baselineskip}

The above implementation allows constraints to be applied in terms of their natural data representation. For example, to apply mean constraints to a product measure, we would first break the product measure into its component discrete measures with ``unpack'', use ``mean'' to impose a mean on the relevant 1D discrete measures, and then use ``pack'' to build the new product measure.
Since constraints are applied on measure objects, but a \verb!costFunction! requires a parameter list \verb!param! as input, ``flatten'' must be used to convert measure objects to a list of weights and positions. After the optimizer provides the new \verb!param! values, ``unflatten'' then converts the modified list of weights and positions into new measure objects.

It is worthwhile to note that the OUQ algorithm presented in the remainder of this subsection is independent of the specific implementation of the measure data objects.
As shown in Figure \ref{fig:OptimizerLoop}, a list of \verb!params! and a \verb!cost! is sufficient to exchange information between portions of the optimization loop.  A more complex object hierarchy is required only within the \verb!costFunction!, and in the application of the non-bounds constraints (which utilize measure object methods).
By providing the data object hierarchy with ``flatten'' and ``unflatten'' methods, the measure data objects can be decoupling from the rest of the algorithm.  All bounds constraints are directly applied to the cost function, and act on a single variable (\emph{i.e.}\ a member of the \verb!params! list), while all non-bounds constraints are applied within the constraint function, which takes \verb!params! as both input and output (as shown in the text below).
Thus, the product measure hierarchy that is chosen to best provide support for the model, whether
it be a convex combination of Dirac masses, or measures composed of another basis such as Gaussians, has no effect on the implementation of the remainder of the OUQ algorithm code.

More precisely, since measures are handled as data objects, this framework can naturally be extended to Gaussians by adding covariance matrices as
data object variables and by estimating integral moments equations (with a Monte Carlo method for instance) instead of using the equations given below. Indeed, \emph{mystic} also provides Gaussian measure objects which can be selected for OUQ optimizations.

\vspace{\baselineskip}

In the basic situation in which Dirac measures are the basis of the 1D measures (as opposed to, say, Gaussians), the optimizer can discover the weights and positions of the Diracs that maximize $\mu[H = 0]$.  Note that, when $\mu$ is a product of convex combinations of Dirac measures, \emph{i.e.}
\begin{equation}
  \label{eq:1dSupport}
	\mu = \left( \sum_{i = 1}^{N_{x}} w_{x_{i}} \delta_{x_{i}} \right) \otimes \left( \sum_{j = 1}^{N_{y}} w_{y_{j}} \delta_{y_{j}} \right) \otimes \left( \sum_{k = 1}^{N_{z}} w_{z_{k}} \delta_{z_{k}} \right),
\end{equation}
the $\mu$-probability of failure $\mu[H = 0]$ and the mean value of the response function $\E_{\mu}[H]$ can be easily calculated as follows:
\begin{equation}
  \label{eq:CostFunction}
	\mu[H = 0] = \sum_{i,j,k} w_{x_{i}} w_{y_{j}} w_{z_{k}} \mathbbm{1}[H(x_{i}, y_{j}, z_{k}) = 0],
\end{equation}
\begin{equation}
  \label{eq:3dExpectation}
	\E_{\mu}[H] = \sum_{i,j,k} w_{x_{i}} w_{y_{j}} w_{z_{k}} H(x_{i}, y_{j}, z_{k}).
\end{equation}

\section{A motivating physical example and associated results}
\label{sec:Example}

In this section, we discuss the results of numerical implementation of OUQ algorithms for a simple surrogate model for hypervelocity impact.  In particular, we compute the upper ($\mathcal{U}(\mathcal{A})$)
 bound on the probability of failure for various sets of assumptions $\mathcal{A}$, and also consider a simple example of the experimental selection problem.

The physical system of interest is one in which a 400C steel ball of diameter $D_{\mathrm{p}} = 1.778 \, \mathrm{mm}$ impacts a 440C steel plate of thickness $h$ (expressed in $\mathrm{mm}$) at speed $v$ (expressed in $\mathrm{km \cdot s}^{-1}$) at obliquity $\theta$ from the plate normal.  The physical experiments are performed at the California Institute of Technology SPHIR (Small Particle Hypervelocity Impact Range) facility.  A simple surrogate model was developed to calculate the perforation area (in $\mathrm{mm}^{2}$) caused by this impact scenario. The surrogate response function is as follows:
\begin{equation}
	\label{eq:PSAAP_SPHIR_surr}
	H(h, \theta, v) = K \left( \frac{h}{D_{\mathrm{p}}} \right)^{p} (\cos \theta)^{u} \left( \tanh \left( \frac{v}{v_{\mathrm{bl}}} - 1 \right) \right)_{+}^{m},
\end{equation}
where the \emph{ballistic limit velocity} (the speed below which no perforation area is caused) is given by
\begin{equation}
	\label{eq:ballistic_limit}
	v_{\mathrm{bl}} := H_{0} \left( \frac{h / \mathrm{mm}}{(\cos \theta)^{n}} \right)^{s}.
\end{equation}
The seven quantities $H_{0}$, $s$, $n$, $K$, $p$, $u$ and $m$ are fitting parameters that have been chosen to minimize the least-squares error between the surrogate and a set of 56 experimental data points;  they take the values
\begin{align*}
	H_{0} &= 0.5794 \, \mathrm{km \cdot s}^{-1}, & s &= 1.4004, & n &= 0.4482, & K &= 10.3936 \, \mathrm{mm^{2}}, \\
	p &= 0.4757, & u &= 1.0275, & m &= 0.4682. &&
\end{align*}

For the remainder of this section, the surrogate response function, $H$, will be taken as given and fixed, and its behaviour will be investigated on the input parameter range
\begin{subequations}
	\begin{align}
		\label{eq:PSAAP_SPHIR_range}
		(h, \theta, v) \in \mathcal{X} & := \mathcal{X}_{1} \times \mathcal{X}_{2} \times \mathcal{X}_{3}, \\
		h \in \mathcal{X}_{1} & := [1.524, 2.667] \, \mathrm{mm} = [65, 105] \, \mathrm{mils}, \\
		\theta \in \mathcal{X}_{2} & := [0, \tfrac{\pi}{6}], \\
		v \in \mathcal{X}_{3} & := [2.1, 2.8] \, \mathrm{km \cdot s}^{-1}.
	\end{align}
\end{subequations}
We will measure lengths in both  $\mathrm{mm}$ and $\mathrm{mils}$ (recall that $1\,\mathrm{mm}=0.0254\,\mathrm{mils}$).
Since $H$ has been fixed, the optimizations that follow are all optimizations with respect to the unknown random distribution $\P \in \mathcal{M}(\mathcal{X})$ of the inputs of $H$.  We adopt the ``gunner's perspective'' that the failure event is non-perforation, \emph{i.e.}\ the event $[H = 0]$.

%%\subsection{Computational Examples}

%% 2010-08-04:  TJS made changes
%% 2010-08-10:  MMM small edits

%%\begin{eg}[Computation of optimal bounds on the probability of non-perforation]
%%	\label{eg:PSAAP_support_collapse}

\paragraph{Computation of optimal bounds on the probability of failure.} We assume that the impact velocity, impact obliquity and plate thickness are independent random variables, and that the mean perforation area must lie in a prescribed range $[m_{1}, m_{2}] := [5.5, 7.5] \, \mathrm{mm}^{2}$.  Therefore, the corresponding admissible set for the OUQ problem is
	\begin{equation}
	\label{eq:PSAAP_SPHIR_Admissible}
		\mathcal{A} := \left\{ (H, \mu) \,\middle|\,
		\begin{matrix}
			H \text{ given by \eqref{eq:PSAAP_SPHIR_surr},} \\
			\mu = \mu_{1} \otimes \mu_{2} \otimes \mu_{3}, \\
			m_{1} \leq \mathbb{E}_{\mu}[H] \leq m_{2}
		\end{matrix}
		\right\}.
	\end{equation}
\COMMENT{
	By way of contrast, an admissible set that corresponds to this problem and the assumptions of McDiarmid's inequality is
	\begin{equation}
	\label{eq:PSAAP_SPHIR_Admissible_McD}
		\mathcal{A}_{\mathrm{McD}} :=
		\left\{ (f, \mu) \,\middle|\,
		\begin{matrix}
			\mu = \mu_{1} \otimes \mu_{2} \otimes \mu_{3}, \\
			m_{1} \leq \mathbb{E}_{\mu}[f] \leq m_{2}, \\
			\operatorname{Osc}_{i}(f) \leq \operatorname{Osc}_{i}(H) \text{ for } i = 1, 2, 3
		\end{matrix}
		\right\}.
	\end{equation}
We refer to equation \eqref{eq:Osc} for the definition of $\operatorname{Osc}_{i}(f)$ and to equation \eqref{eq:MDassumptions} for measurability issues.
	This example serves to illustrate the extent to which McDiarmid's inequality is not sharp.  Application of McDiarmid's inequality to the above situation yields the bound
	\[
		\P[H = 0] \leq \mathcal{U}(\mathcal{A}_{\mathrm{McD}}) \leq \exp \left( - \frac{2 m_{1}^{2}}{\sum_{i = 1}^{3} \operatorname{Osc}_{i}(H)^{2}} \right) = 66.4\%.
	\]
	The ``optimal McDiarmid inequality'' of Theorem \ref{thm:m3} and Remark \ref{rmk:baby_measure_indexing} instead provide the upper bound
	\[
		\P[H = 0] \leq \mathcal{U}(\mathcal{A}_{\mathrm{McD}}) = 43.7\%.	
	\]
}	
	With access to $H$, bounds on the probability of non-perforation can be calculated.  In order to find the optimal upper bound, $\mathcal{U}(\mathcal{A})$, it is sufficient to search among those measures $\mu$ whose marginal distributions on each of the three input parameter ranges have support on at most two points \cite{OUQ:2010}.  That is, $\mathcal{U}(\mathcal{A}) = \mathcal{U}(\mathcal{A}_{\Delta})$,	where the reduced feasible set $\mathcal{A}_{\Delta}$ is given by
	\begin{equation}
	\label{eq:PSAAP_SPHIR_Admissible_Reduced}
		\mathcal{A}_{\Delta} := \left\{ (H, \mu) \,\middle|\,
		\begin{matrix}
			H \text{ given by \eqref{eq:PSAAP_SPHIR_surr},} \\
			\mu = \mu_{1} \otimes \mu_{2} \otimes \mu_{3}, \\
			\mu_{i} \in \Delta_{1}(\mathcal{X}_{i}) \text{ for } i = 1, 2, 3, \\
			m_{1} \leq \mathbb{E}_{\mu}[H] \leq m_{2}
		\end{matrix}
		\right\}.
	\end{equation}
	A numerical optimization over this finite-dimensional reduced feasible set $\mathcal{A}_{\Delta}$ using a genetic algorithm global optimization routine in the \emph{mystic} framework yields the following optimal upper bound on the probability of non-perforation:
	\[
		\P[H = 0] \leq \mathcal{U}(\mathcal{A}) = \mathcal{U}(\mathcal{A}_{\Delta}) \numeq 37.9\%.
	\]
Observe that $\P[H = 0] \leq \mathcal{U}(\mathcal{A}) = \mathcal{U}(\mathcal{A}_{\Delta})$ is a theorem whereas the $\mathcal{U}(\mathcal{A}_{\Delta}) \numeq 37.9\%$ is the output of an algorithm (in this case, a Differential Evolution Algorithm  implemented in the \emph{mystic} framework, see Section \ref{sec:Code}). In particular, its validity is correlated with the efficiency of the specific algorithm. We have introduced the symbol $\numeq$ to emphasize the distinction between mathematical equalities and numerical outputs.

 Although we don't have a theorem associated with the convergence of the numerical optimization algorithm,   we have a robust control over its efficiency because  it is applied to the finite dimensional problem $\mathcal{U}(\mathcal{A}_{\Delta})$ instead of the infinite optimization problem associated with $\mathcal{U}(\mathcal{A})$.

\begin{figure}[tp]
	\begin{center}
		\subfigure[support points at iteration 0]{
			\includegraphics[width=0.45\textwidth]{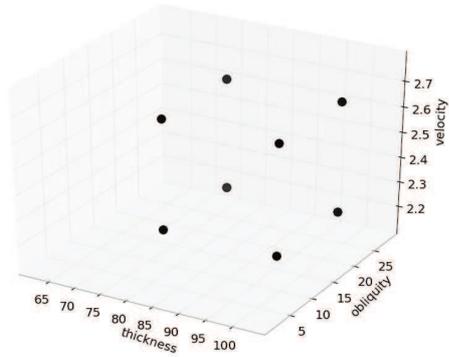}
		}
		\subfigure[support points at iteration 150]{
			\includegraphics[width=0.45\textwidth]{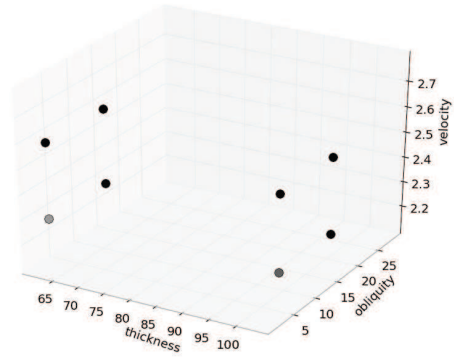}
		}
		\subfigure[support points at iteration 200]{
			\includegraphics[width=0.45\textwidth]{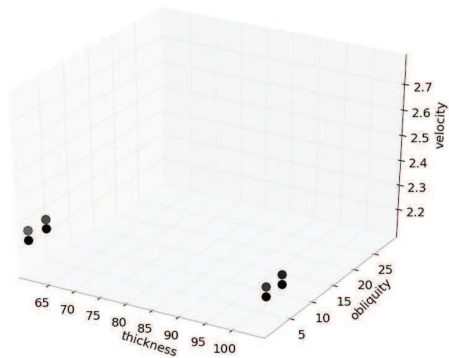}
		}
		\subfigure[support points at iteration 1000]{
			\includegraphics[width=0.45\textwidth]{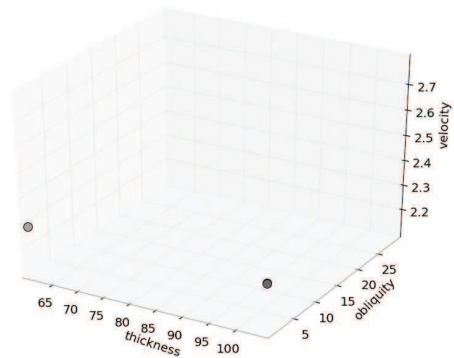}
		}
		\caption{
For  $\# \mathrm{supp}(\mu_{i}) \le 2, \, i = 1, 2, 3$,  the maximizers collapse to two-point support. Velocity and obliquity marginals each collapse to a single Dirac mass, while the plate thickness marginal collapses to have support on the extremes of its range.  Note the perhaps surprising result that the probability of non-perforation is maximized by a distribution supported on the minimal, not maximal, impact obliquity.}
		\label{fig:CollapseSupport2}
	\end{center}
\end{figure}

\begin{figure}[tp]
	\begin{center}
		\subfigure[convergence for thickness]{
			\includegraphics[width=0.40\textwidth]{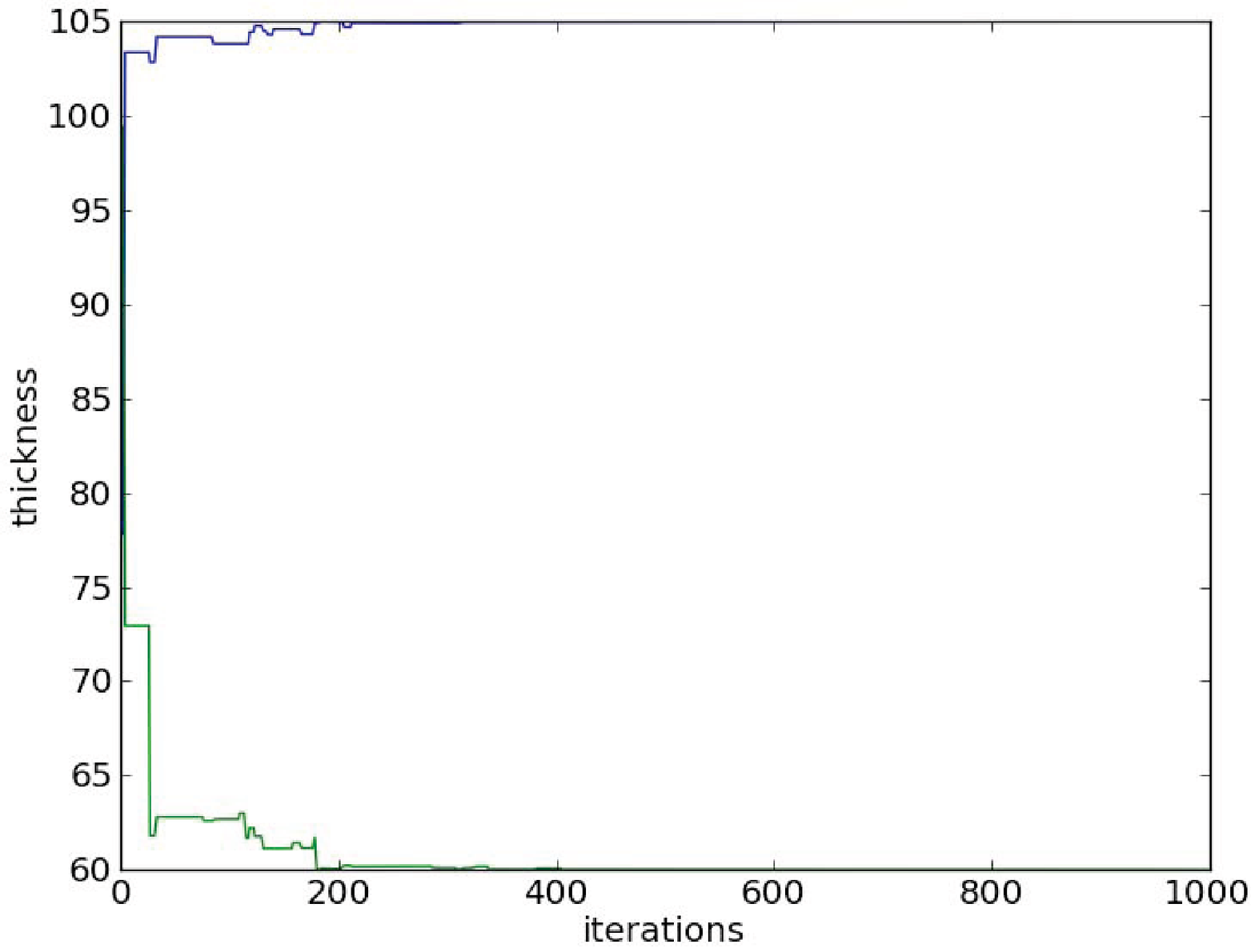}
		}
		\subfigure[convergence for thickness weight]{
			\includegraphics[width=0.40\textwidth]{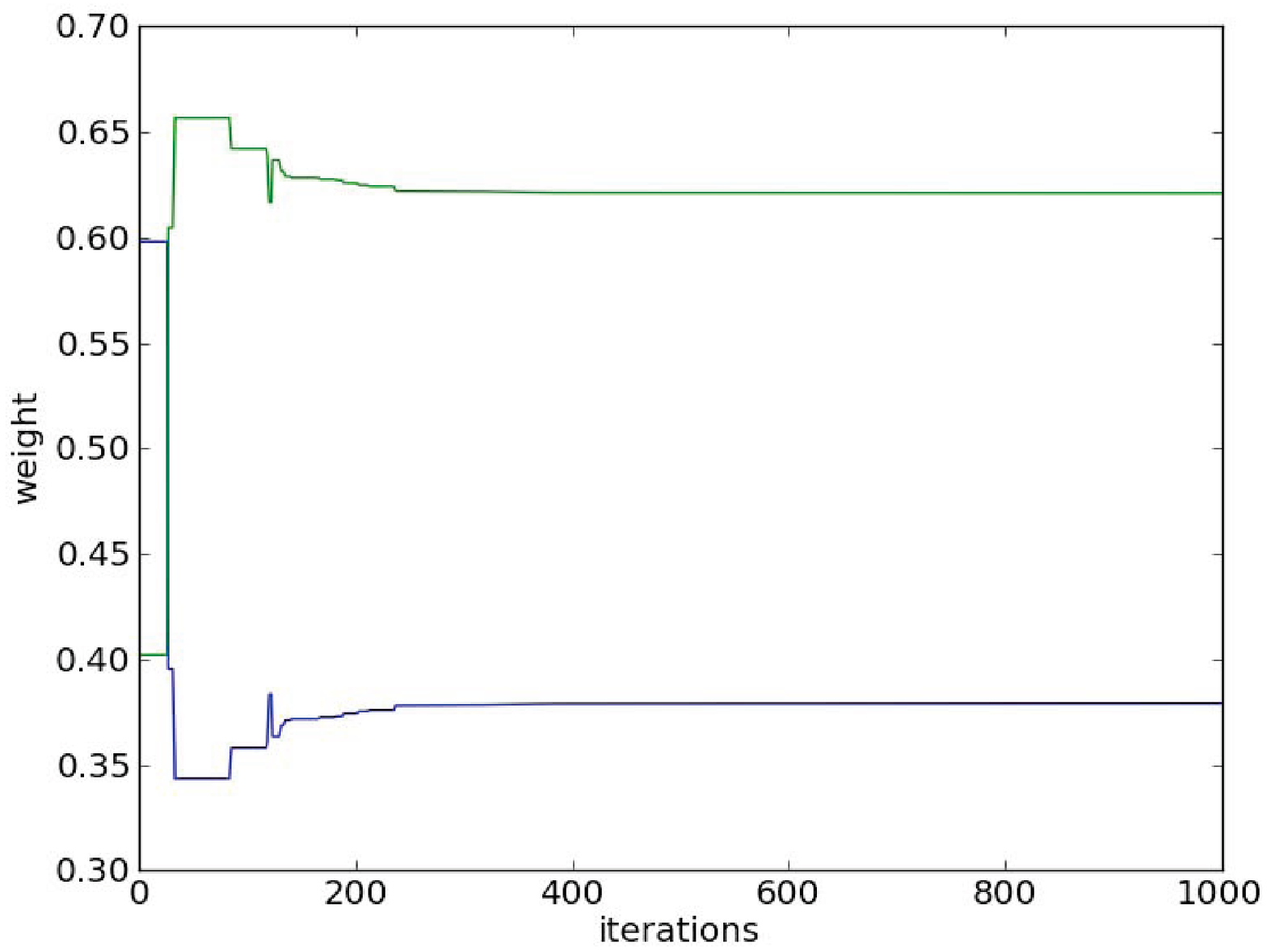}
		}
		\subfigure[convergence for obliquity]{
			\includegraphics[width=0.40\textwidth]{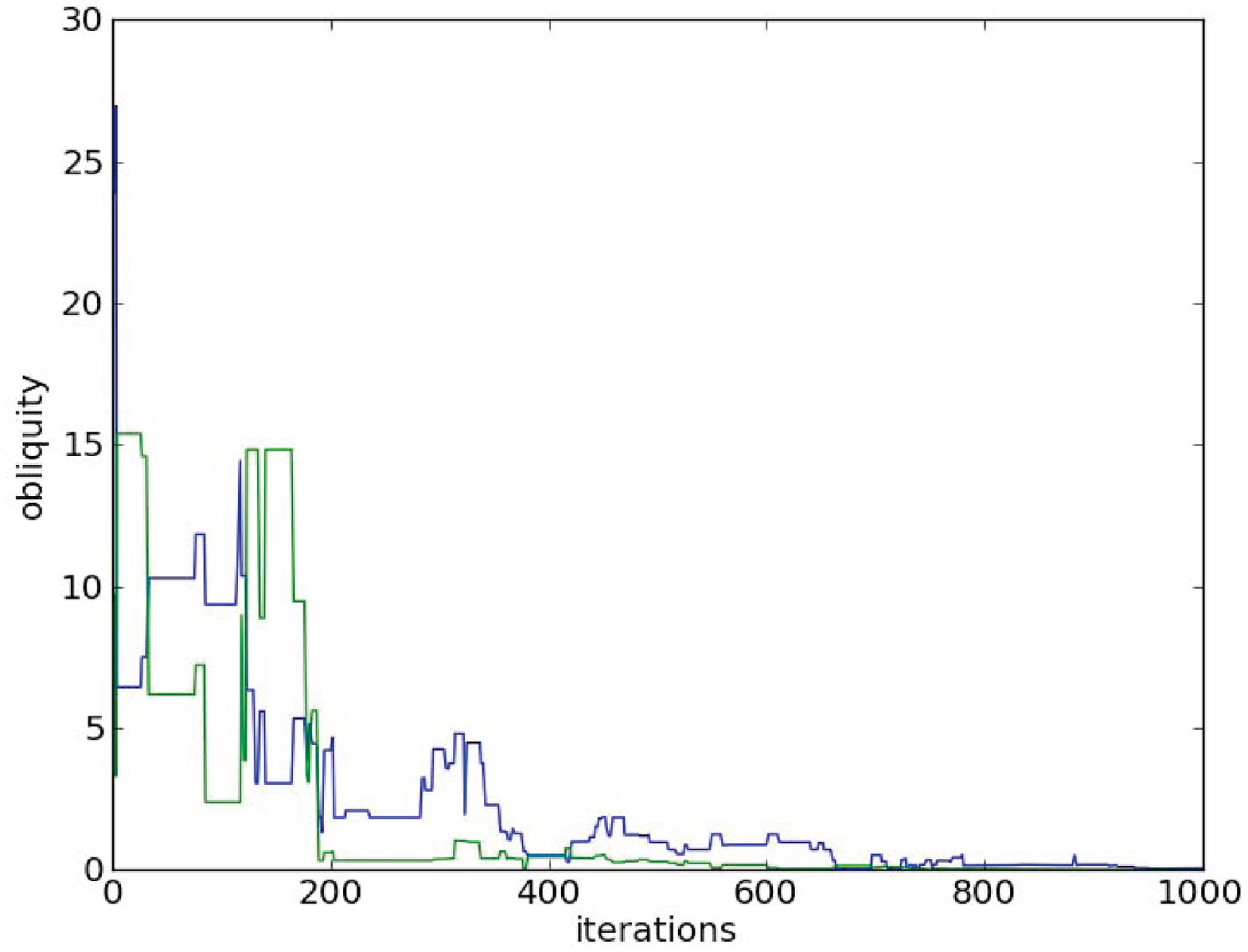}
		}
		\subfigure[convergence for obliquity weight]{
			\includegraphics[width=0.40\textwidth]{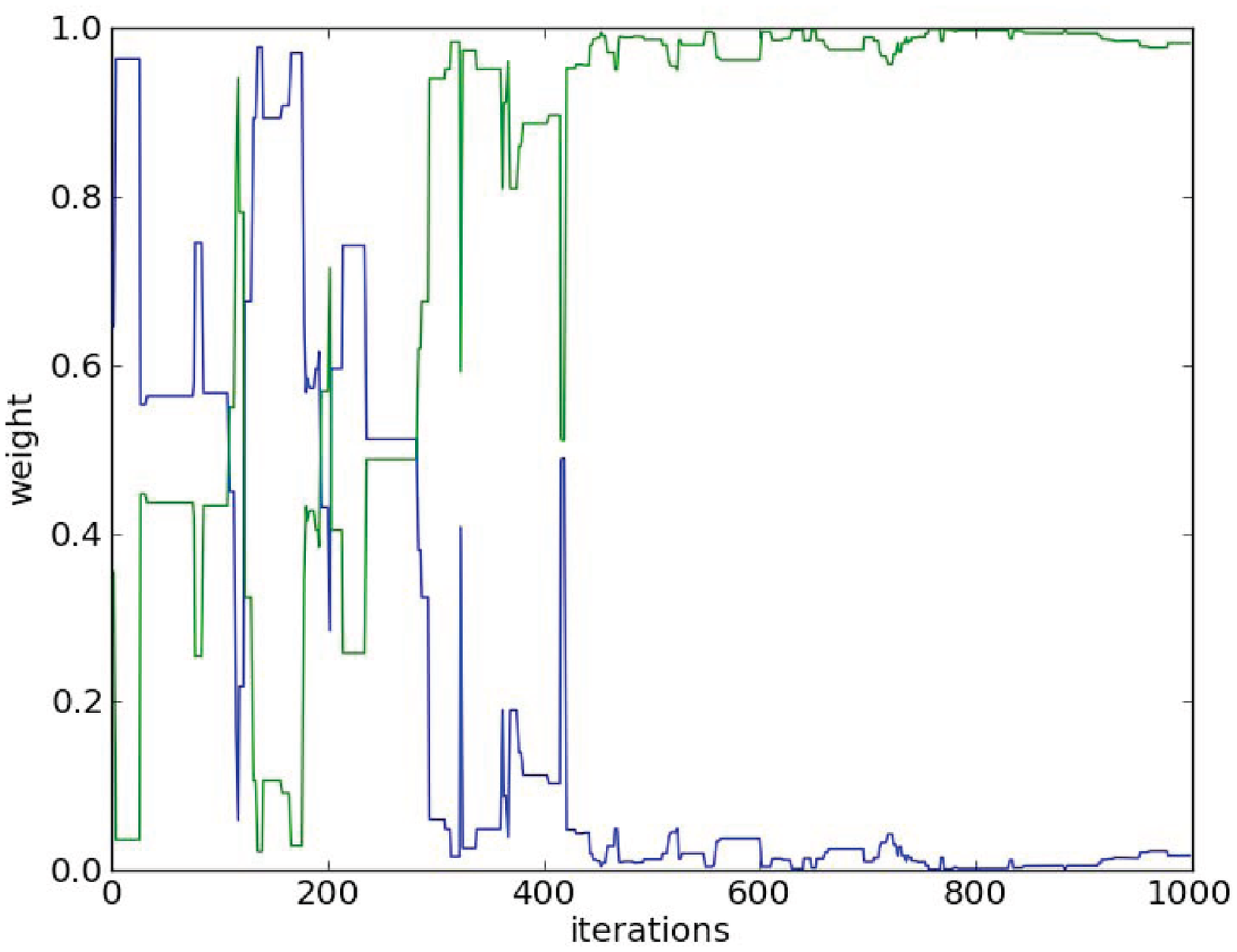}
		}
		\subfigure[convergence for velocity]{
			\includegraphics[width=0.40\textwidth]{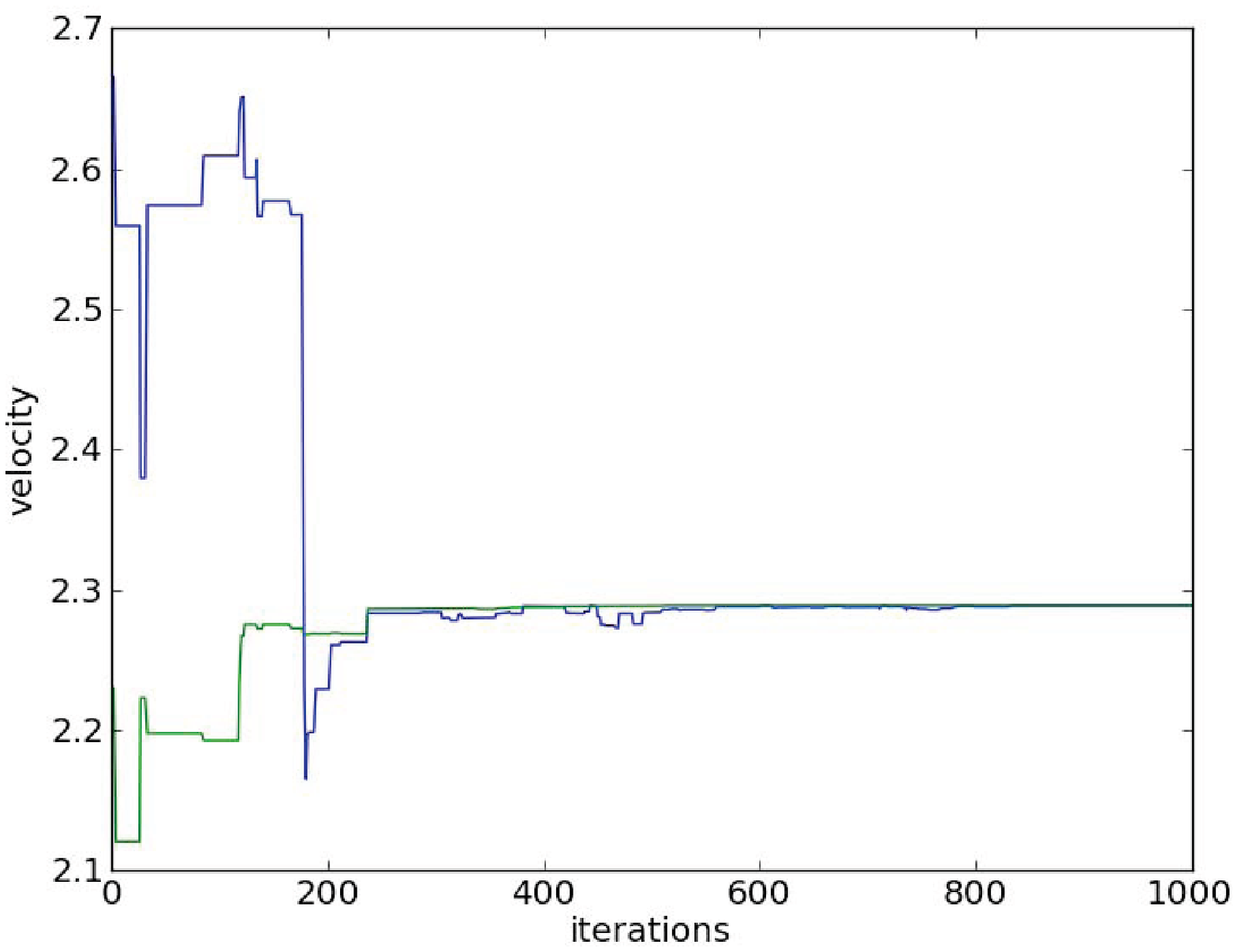}
		}
		\subfigure[convergence for velocity weight]{
			\includegraphics[width=0.40\textwidth]{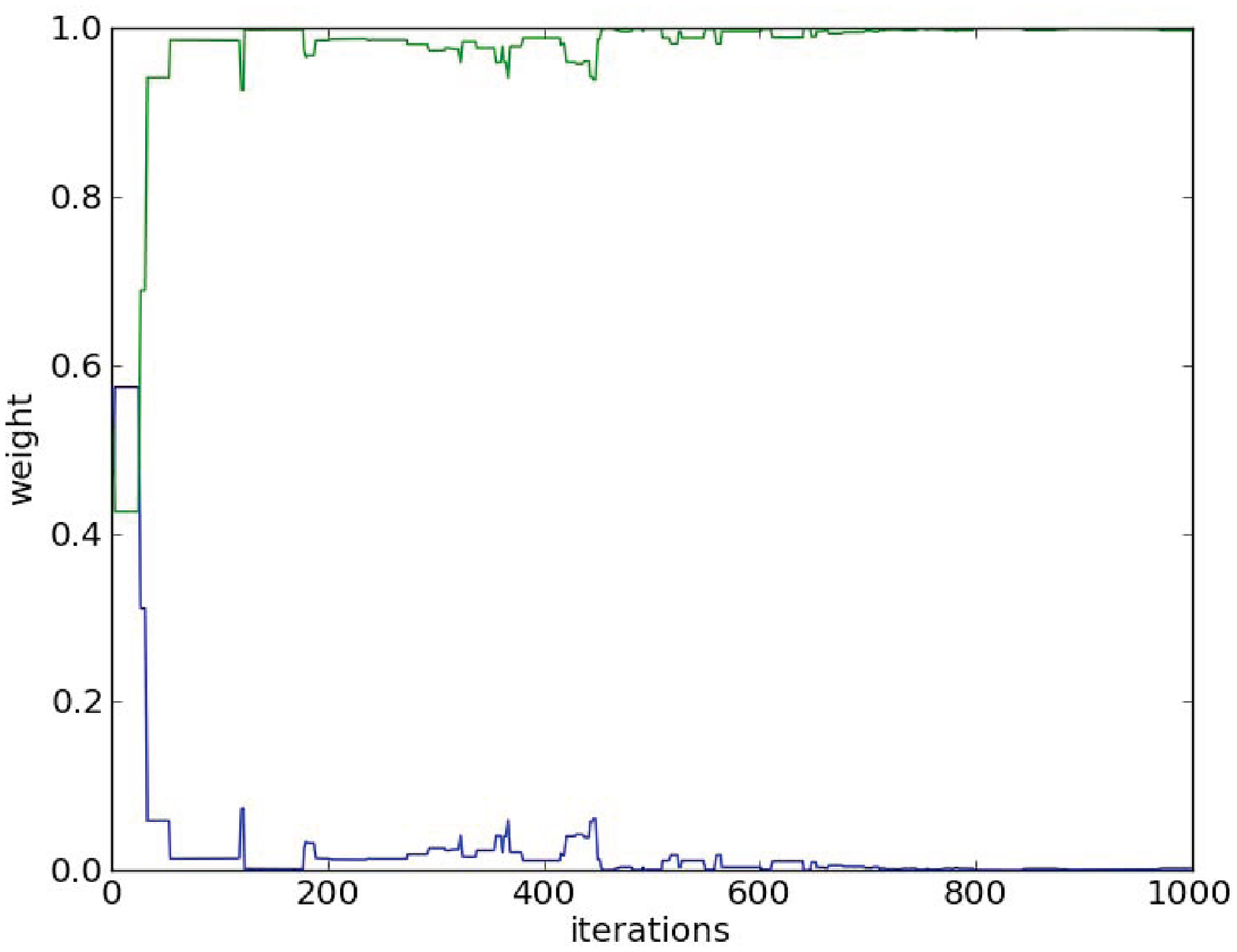}
		}
		\caption{Time evolution of the maximizers for $\# \mathrm{supp}(\mu_{i}) \leq 2$ for $i = 1, 2, 3$, as optimized by \emph{mystic}. Thickness quickly converges to the extremes of its range, with a weight of 0.621 at $60 \, \mathrm{mils}$ and a weight of 0.379 at $105 \, \mathrm{mils}$. The degeneracy in obliquity at 0 causes the fluctuations seen in the convergence of obliquity weight.  Similarly, velocity converges to a single support point at 2.289 $\mathrm{km \cdot s}^{-1}$, the ballistic limit velocity for thickness $105 \, \mathrm{mils}$ and obliquity $0$ (see \eqref{eq:ballistic_limit}).}
% ballistic limit at (h,a) = (105,0): 2.2886591712123701.  Mystic converges to 2.2886591712123701.
% NOTE: do we need to discuss why we have vbl at h,a = 105,0 -- and not at h,a = 60,0?
		\label{fig:CollapseConverge2}
	\end{center}
\end{figure}

For  $\# \mathrm{supp}(\mu_{i}) \le 2, \, i = 1, 2, 3$ (where $\# \mathrm{supp}(\mu_{i})$ is the number of points forming the support of $\mu_i$),  Figure \ref{fig:CollapseSupport2} shows that numerical simulations collapse to
 two-point support.  Indeed, even when a wider search is performed (\emph{i.e.}\ over measures $\mu \in \bigotimes_{i = 1}^{3} \Delta_{k}(\mathcal{X}_{i})$ for $k > 1$), it is observed that the calculated maximizers for these problems maintain two-point support:  the velocity and obliquity marginals each collapse to a single Dirac mass, and the plate thickness marginal collapses to have support on the two extremes of its range.  As expected, optimization over a larger search space is more computationally intensive and takes longer to perform.
%% This observation suggests that the extreme points of the reduced  OUQ problems are, in some sense, attractors --- this point will be revisited in the next subsection.
We also refer to Figure \ref{fig:CollapseConverge2}
%%, \ref{fig:CollapseConverge3} and \ref{fig:CollapseConverge5}
for plots of the locations and weights of the Dirac masses forming each marginal $\mu_i$ as  functions of the number of iterations. Note that the lines for \emph{thickness} and \emph{thickness weight} are of the same color if they correspond to the same support point for the measure.
%%In particular, Figure \ref{fig:CollapseConverge5} shows that at iteration the \emph{thickness} support point at $62.5\, \mathrm{mils}$ (shown in Figure \ref{fig:CollapseSupport5}) has zero weight.

\section{Implementation of the physical example in \emph{mystic}}
\label{sec:Code}

\paragraph{The OUQ outer loop.} As posed above, an OUQ problem at the high-level is a global optimization of a cost function that satisfies a set of constraints.
The optimization of the example presented in Section \ref{sec:Example} is performed in \emph{mystic} using the differential evolution algorithm of Price \& Storn \cite{PriceStornLampinen:2005, StornPrice:1997}.  Differential evolution is a form of genetic algorithm that generates \verb!npop! trial solutions each iteration (here \verb!npop! is called the ``trial population size''), and uses a ``mutation strategy'' to modify trial solutions in attempt to produce a better solution.  Beyond selecting the mutation strategy, two parameters control the amount of mutation that occurs at each iteration: \verb!CrossProbability! and \verb!ScalingFactor!.  \verb!CrossProbability! is the chance that a mutation will occur, while \verb!ScalingFactor! is the relative size of the resulting mutation.

For the problem at hand, the optimizer is configured to use a trial population size of \verb!npop = 40!, \verb!CrossProbability = 0.9!, \verb!ScalingFactor = 0.9!, and mutation strategy = \verb!Best1Exp! \cite{McKernsHungAivazis:2009}. Hence, there are forty trial solutions generated each iteration, where a \verb!trialSolution! is equal to the current \verb!bestSolution! plus the scaled difference of two of the other randomly selected trial solution \verb!candidates!. The trial solution is altered (\emph{i.e.}\ ``undergoes a mutation'') 90\% of the time.
The optimizer begins by selecting initial parameter values at random from a uniform distribution over the bounds, and the following code demonstrates how the strategy is used to generate a new parameter value at each successive iteration.
\begin{verbatim}
if random.random() >= CrossProbability:
    trialSolution = bestSolution
else:
    trialSolution = bestSolution+ScalingFactor*(candidate1 - candidate2)
\end{verbatim}
This is the standard implementation of the \verb!Best1Exp! mutation strategy for Differential Evolution, as detailed in   \cite{PriceStornLampinen:2005} (we also refer to \cite{PriceStornLampinen:2005} for a precise definition of candidate1 and candidate2).

The trial parameter set \verb!param! is the collection of \verb!trialSolutions!. After generating \verb!param!, \emph{mystic} evaluates the \verb!costFunction! at each iteration to generate the goodness of fit, and then checks for convergence.
For OUQ, the fit parameters \verb!param! are the positions and weights of the 1D discrete measure support along each axis of the hypercube. For example, if we use two Dirac masses as support in each direction, the fit parameters used will be:
\begin{verbatim}
param = [wx1, wx2, x1, x2, wy1, wy2, y1, y2, wz1, wz2, z1, z2]
\end{verbatim}

\emph{Mystic} provides the ability to impose both bounds constraints and non-bounds constraints in a general way.
A constraint function \verb!constrain! is applied at every optimizer iteration, forcing all random variables generated as a trial parameter set to satisfy the non-bounds constraints before the cost function is evaluated.
\begin{verbatim}
param = constrain(param)
cost = costFunction(param)
\end{verbatim}
Bounds constraints are enforced on the random variables at each iteration by imposing an infinite potential well on the cost function, using the keyword \verb!SetStrictRanges! \cite{McKernsHungAivazis:2009}.

\verb!ChangeOverGeneration(tolerance=1e-4, generations=10)! was selected as the termination condition \cite{McKernsHungAivazis:2009}. Thus convergence is defined as when the change in cost over ten iterations is less than 0.0001.
If the convergence criteria are met, the \verb!param! are returned as calculated global maximizers \verb!opt_param!, with the resulting calculated maximum at \verb!opt_cost = costFunction(opt_param)!.  The optimizer thus provides \verb!opt_cost! as the numerical calculation of $\mathcal{U}(\mathcal{A})$.

For the example in Section \ref{sec:Example}, the cost function is \eqref{eq:CostFunction}, where $h=x$, $\theta=y$, and $v=z$. Similarly for $h,\theta,v$, the constraints imposed by \verb!constrain! are provided by \eqref{eq:3dExpectation} and
\[
	m_{1} \leq \mathbb{E}_{\mu}[H] \leq m_{2}, \text{where } [m_{1}, m_{2}] = [5.5, 7.5] \, \mathrm{mm}^{2},
\]
\[
  \sum_{i = 1}^{N_{h}} w_{h_{i}} = 1, \quad
  \sum_{j = 1}^{N_{\theta}} w_{\theta_{j}} = 1, \quad
  \sum_{k = 1}^{N_{v}} w_{v_{k}} = 1.
\]
The global optimization is also constrained by bounds constraints provided by \eqref{eq:PSAAP_SPHIR_range} and
\[
	w_{h_{i}} \in [0, 1] \text{ for each } i \in \{ 1, \dots, N_{h}\},
\]
\[
	w_{\theta_{j}} \in [0, 1] \text{ for each } j \in \{ 1, \dots, N_{\theta} \},
\]
\[
	w_{v_{k}} \in [0, 1] \text{ for each } k \in \{ 1, \dots, N_{v} \}.
\]

\COMMENT{
    We generate the parameter constraint function for the given m, d, and (Nx,Ny,Nz).
      * A factory method is used to generate a constraint function \verb!constrain(param)!
    We generate the maximizing function for the given theta.
      * A factory method is used to generate the \verb!model(product_measure)!
    After solving, we unflatten and pack the calculated maximizers, producing the optimal product measure.
    Additionaly, expectation E is computed for optimal support, E(opt_product_measure).
}

The constraint function \verb!constrain! is used to ensure the trial parameters \verb!param! generated by the optimizer satisfy the non-bounds constraints. The \verb!param! are first unflattened to $3$ one-dimensional discrete measures, and each measure is checked for weight normalization (\emph{i.e.}\ that the total mass of the measure is $1$);  each measure for which the sum of the weights does not equal $1$ is re-normalized to have total weight $1$.  The $3$ one-dimensional discrete measures are then packed into a single three-dimensional product measure.  The product measure is then checked for compliance with the mean constraint.

The mean perforation area constraint provides that the expectation of the product measure is in $[m - d, m + d]:= [m_1, m_2]$ (where $m$ is the target mean, and $d$ defines some acceptable deviation from the target mean).  If this constraint is not satisfied,  then a new perforation area is imposed in $[m - d, m + d]$  is imposed by using the \emph{expect} method provided by the measure object, and described below (imposing a new value for the mean preformation area is an optimization as described by the OUQ inner loop).

 The product measure is then unpacked and flattened to produce the resulting trial parameters.  The implementation of \verb!constrain! is as follows: %NOTE: '!=' is '$\not=$'
\begin{verbatim}
x_measure, y_measure, z_measure = unflatten(param)

if sum(x_measure.weights) != 1.0:
    x_measure.normalize()
if sum(y_measure.weights) != 1.0:
    y_measure.normalize()
if sum(z_measure.weights) != 1.0:
    z_measure.normalize()

product_measure = pack((x_measure, y_measure, z_measure))

if product_measure.expect(H) > (m+d) or product_measure.expect(H) < (m-d):
    product_measure.expect = m +/- d

param = flatten(unpack(product_measure))
\end{verbatim}

\paragraph{The OUQ inner loop.} The reader may have noticed that satisfying the constraints provided by \eqref{eq:3dExpectation} can also be formulated as an optimization problem.
The goal is to obtain a new set of support points that satisfy the condition $E[H] \in [m_1, m_2]$ (where $[m_1, m_2] = [m - d, m + d]$).
The optimization is again performed in \emph{mystic} using differential evolution.
For the inner loop, the optimizer is configured to use a trial population size of 20, \verb! CrossProbability = 0.9!, \verb!ScalingFactor = 0.9!, and mutation strategy = \verb!Best1Exp!.
The initial parameter values and trial parameter values are generated as above. Bounds constraints are imposed on the random variables using \verb!SetStrictRanges!, as above.
The cost function for the inner loop is the least squared difference of $\mathbb{E}_{\mu}[H]$ and the target mean $m$.
\begin{verbatim}
product_measure = pack(unflatten(param))
cost = (product_measure.expect(H) - m)**2
\end{verbatim}
The termination condition used for the inner loop is \verb!VTR(tolerance = d^2)! \cite{McKernsHungAivazis:2009}, which defines convergence as having occured when the change in the cost from the last iteration is less than the prescribed \verb!tolerance!.  Therefore, if the convergence criteria are met, the resulting maximizers yield a \verb!product_measure! where $\mathbb{E}_{\mu}[H] \in [m - d, m + d]$.

\section*{Acknowledgements}

The authors gratefully acknowledge  portions of this work supported by the Department of Energy National Nuclear Security Administration under Award Number DE-FC52-08NA28613 and by the National Science Foundation under Award Number DMR-0520547.

%GATHER{./refs.bib}
\bibliographystyle{plain}
\bibliography{./refs}

\def\polhk#1{\setbox0=\hbox{#1}{\ooalign{\hidewidth
  \lower1.5ex\hbox{`}\hidewidth\crcr\unhbox0}}}
\begin{thebibliography}{1}

\bibitem{McKernsHungAivazis:2009}
M.~McKerns, P.~Hung, and M.~Aivazis.
\newblock Mystic: A simple model-independent inversion framework, 2009.
\newblock \url{http://dev.danse.us/trac/mystic}.

\bibitem{OUQ:2010}
H.~Owhadi, T.~J. Sullivan, M.~McKerns, M.~Ortiz, and C.~Scovel.
\newblock Optimal uncertainty quantification.
\newblock {\em SIAM Review}, Submitted, 2010.
\newblock \url{http://arxiv.org/pdf/1009.0679}.

\bibitem{PriceStornLampinen:2005}
K.~V. Price, R.~M. Storn, and J.~A. Lampinen.
\newblock {\em Differential {E}volution}.
\newblock Natural Computing Series. Springer-Verlag, Berlin, 2005.
\newblock A practical approach to global optimization, With 1 CD-ROM (Windows,
  Macintosh and UNIX).

\bibitem{StornPrice:1997}
R.~M. Storn and K.~V. Price.
\newblock Differential evolution --- a simple and efficient heuristic for
  global optimization over continuous spaces.
\newblock {\em J. Global Optim.}, 11(4):341--359, 1997.
\newblock \url{http://dx.doi.org/10.1023/A:1008202821328}.

\end{thebibliography}

\end{document}